\newif\iflinenumbers
\linenumbersfalse %

\documentclass[10pt,oneside,notitlepage,abstracton,a4paper]{article}

\usepackage{epsfig,scrlayer-scrpage}
\usepackage{amsmath}
\usepackage{makecell}
\usepackage{graphicx}
\usepackage{booktabs}
\usepackage{arydshln}
\usepackage{setspace} %
\usepackage[utf8]{inputenc}
\usepackage[T1]{fontenc}

  \date{\normalsize \today}

\usepackage{chngcntr}

\usepackage[bottom]{footmisc}

\usepackage{titlesec}
\titleformat*{\section}{\large\bfseries}
\titleformat*{\subsection}{\normalsize\bfseries}

\iflinenumbers
  \usepackage{lineno}
  \linenumbers
\fi

\usepackage[affil-it]{authblk}

\usepackage{etoolbox} %
\makeatletter %
\patchcmd{\@maketitle}{\LARGE \@title}{\fontsize{16}{19.2}\selectfont\@title}{}{}
\makeatother
\usepackage[affil-it]{authblk}

\usepackage{geometry}
\newcommand{\imagepath}{.}

\usepackage{xcolor}
\definecolor{DESYcyan}{RGB}{0,166,235}
\definecolor{DESYorange}{RGB}{242,142,0}
\definecolor{DESYgray}{RGB}{119,119,119}
\definecolor{bjetpurple}{RGB}{117,112,179}

\usepackage[compat=1.1.0]{tikz-feynman}
\usepackage{contour}
\usetikzlibrary{arrows,shapes,positioning}
\usetikzlibrary{decorations.markings}
\usetikzlibrary{decorations.pathmorphing}
\usetikzlibrary{decorations.pathreplacing}
\usetikzlibrary{patterns}
\usetikzlibrary{plotmarks}
\usetikzlibrary{shadows}

\usepackage[margin=8mm,font=small,labelfont=bf,format=plain]{caption}
\usepackage[margin=8mm,font=small,labelfont=bf,format=plain]{subcaption}

\usepackage{multirow} %

\newcommand{\subfigref}[1]{({\protect\subref{#1}})}

\usepackage[
  bookmarks,                   %
  bookmarksopenlevel=1,        %
  bookmarksnumbered=true,      %
  pdfstartpage={1},            %
  pdfstartview={FitH},         %
  pdfkeywords={},              %
  pdfsubject={},               %
  pdfcreator={LaTeX with KOMA-Script and hyperref package},
  hyperfootnotes=true,         %
  linkbordercolor={0 1 1},     %
  menubordercolor={0 1 1},     %
  urlbordercolor={1 0 0}       %
]{hyperref}                    %

\usepackage{cleveref}

\usepackage{slashed}

\newcount\colveccount
\newcommand*\colvec[1]{
        \global\colveccount#1
        \begin{pmatrix}
        \colvecnext
}
\def\colvecnext#1{
        #1
        \global\advance\colveccount-1
        \ifnum\colveccount>0
                \\
                \expandafter\colvecnext
        \else
                \end{pmatrix}
        \fi
}

\usepackage{lmodern}
\usepackage{listings}
\usepackage[english]{babel}%
\usepackage{csquotes}%
\usepackage[style=numeric-comp,sorting=none]{biblatex}
\newcommand{\customprintbibliography}{%
  \printbibliography[heading=none]
  \newpage
}

\usepackage{amssymb}

\newcommand{\Lumi}{\mathcal{L}}
\newcommand{\Pol}{\mathcal{P}} %
\newcommand{\PoleM}{\Pol_{\eM}}
\newcommand{\PoleP}{\Pol_{\eP}}

\newcommand{\eP}{e^{+}}
\newcommand{\eM}{e^{-}}

\newcommand{\eML}{\eM_{L}}
\newcommand{\ePR}{\eP_{R}}

\newcommand{\muP}{\mu^{+}}
\newcommand{\muM}{\mu^{-}}

\newcommand{\nubar}{\bar{\nu}}

\newcommand{\Dw}{\Delta w}
\newcommand{\Dc}{\Delta c}
\newcommand{\chisq}{\chi^2}
\newcommand{\chisqshift}{\chisq_{shift}}
\newcommand{\chisqpar}{\chisq_{par}}
\newcommand{\costh}{\cos\theta}
\newcommand{\invab}{\,\text{ab}^{-1}}

\renewcommand{\deg}{^{\circ}}

\title{Isolating systematic effects with beam polarisation at e+e- colliders}
\author[1,2]{Jakob Beyer\thanks{\textit{\footnotesize{Talk presented at the International Workshop on Future Linear Colliders (LCWS2021),}\\ \footnotesize{15-18 March 2021. C21-03-15.1.}}}}
\author[1]{Jenny List}
\affil[1]{Deutsches Elektronen-Synchroton (DESY), Hamburg, Germany}
\affil[2]{Universit\"at Hamburg, Hamburg, Germany}
\date{}

\begin{document}

\maketitle

\begin{abstract}
  Future high-energy $\eP\eM$ colliders will provide some of the most precise tests of the Standard Model. 
  Statistical uncertainties are expected to improve by orders of magnitude over current measurements.
  This provides a new challenge in accurately assessing and minimizing systematic uncertainties. 
  Beam polarisation may hold a unique potential to isolate and determine the size of systematic effects.
  This study aims to set this hypothesis by setting up a combined fit of systematic effect and electroweak observables that can test different configurations of available beam polarisations and luminosities.

  A framework for this fit is developed.
  In a first test it fits the luminosity and polarisation with external constraints to 2-fermion and 4-fermion differential distributions.
  An implementation of a first systematic - the muon acceptance - shows consistent behaviour in this framework.
  This lays the groundwork for testing the interplay of systematic and physical effects in the presence or absence of beam polarisation.
\end{abstract}

\clearpage

\section{Introduction}
Measurements at the SLC demonstrated the physical insight that polarised beams bring \cite{LEP:2003aa}.
Similar advantages from beam polarisation are seen in studies for future colliders  \cite{fujii2019tests}.
The focus in the latter have so far been on the purely physical advantages. 

In addition to physical benefits, beam polarisation may have the potential to also improve on systematic uncertainties. 
This thought bases on the polarisation-independent nature of the majority of systematic effects.
Such systematic effects stay constant when the helicity of the beam is reversed, while the chirality-dependent physical effects change.

A combined fit of the polarised datasets at future colliders can make use of this qualitative difference between physical and systematic effects.
The systematic effects will stand out as global amongst the different polarised datasets if they are explicitly included in the fit.
Resulting from this obvious difference between effects, the parameters of the systematic effects would then not correlate with those of the physical effects.
This uncorrelated behaviour could minimize systematic biases and uncertainties in the physical results.

Here, a framework is set up to study such a combined fit of polarised datasets that explicitly includes systematic effects.
First tests validate the framework for different luminosity and polarisation configurations.
A systematic effect - the $\mu$ acceptance - is parametrised and for the first time included in such a combined fit.
This work lays the foundation for testing the hypothesis that beam polarisation helps to reduce the impact of systematic effects.

\section{A combined fit setup}
An assessment of the impact of beam polarisation on systematic uncertainties at future colliders requires a combined fit of the polarised datasets.
This section describes the framework can perform such fits, the Monte-Carlo distributions used in it, and a first validation test of the framework.

\subsection{A fit framework for polarised datasets}\label{subsec:FitFramework}

A framework is set up that can use all polarised (or unpolarised) distributions for multiple final states at a future collider in a simultaneous fit.
The fit performs a Poissonian log-likelihood maximisation using the \texttt{Minuit2} framework of \texttt{ROOT}.
All polarisations and the luminosity can vary freely, be given a Gaussian constraint, or set to a fixed value.
The helicity flip between polarised datasets is not assumed to be perfect, so that each polarisation of each dataset is a separate parameter.

Proposed future colliders motivate the polarisation and luminosity configurations used in this study \cite{eppsupg2020physics}.
The polarisation scenarios are
\begin{itemize}
  \item an unpolarised scenario ($\PoleM = \PoleP = 0\%$) with the fully luminosity in this unpolarised dataset,
  \item a $\eM$-only polarised scenario with $\PoleM = \pm 80\%$ and $\PoleP=0\%$, and a luminosity sharing of $50:50$ between the two datasets,
  \item a fully polarised scenario with $\PoleM = \pm 80\%$ and $\PoleP = \pm 30\%$, and an optimized luminosity sharing of $45:45:5:5$ where the first two are the opposite-sign and the second two the same-sign datasets.
\end{itemize}
Two luminosity scenarios with a factor 5 difference are tested: $\Lumi = 2\invab$ and $\Lumi = 10\invab$.
All scenarios use realistic Gaussian constraints for each polarisation and luminosity parameter \cite{bartels2009compton,Jelisav_i__2013}.
\begin{equation*}
  \begin{split}
    \frac{\Delta L}{L} &= 3\times10^{-3}\\
    \frac{\Delta \Pol_{non-0}}{\Pol_{non-0}} &= 2.5\times10^{-3} \text{ (for polarised beams)}\\
    \Delta \Pol_{0} &= 2.5\times10^{-3} \text{ (for unpolarised beams)}
  \end{split}
\end{equation*}

\subsection{Processes and distributions}

This work uses differential distributions of di-muon production and semileptonic $W$-pair production at $250\,$GeV.
It uses generator-level event created for the $250\,$GeV ILD production \cite{Berggren_2021}, and are produced using WHIZARD2.8 \cite{Kilian_2011,Moretti:2001zz}.
Previous work within a similar framework for polarisation measurements used a set of processes with two and four fermions in the final state \cite{Beyer:449759}.
The first tests presented here select a subset of those processes that all contain muons in order to test common muon-relevant systematic effects later on.
All distributions are purely on generator level and do not make use of detector simulation or high-level analysis.

The di-muon distributions are one-dimensional distributions of the $\muM$ polar angle in the di-muon rest frame (example in \cref{fig:DistributionExamples_mumu}).
Initial state radiation (ISR) in such events can cause the effective center-mass-energy $\sqrt{s*}$ to change significantly.
Events with strong ISR can land on the $Z$-pole resonance.
Such events are treated separately so that two distinct distributions are used: one with $\sqrt{s*}\in[180,275]\,$GeV (``high-$Q^2$'') and one with $\sqrt{s*}\in[81,101]\,$GeV (``return-to-Z'').
The upper cut in the high-$\sqrt{s*}$ region is higher than $250\,$GeV to account for the beam spectrum.

The semileptonic $W$ pair production distributions describe the $qq\mu\nu\,(q=u,d,s,c,b)$ final state with a three-dimensional angular distribution (example in \cref{fig:DistributionExamples_4f}).
These three angles are the $W^-$ angle in the detector frame and the $\mu$ angles in the rest frame of the leptonically decaying $W$.
Events with negative $\muM$ are used separately from events with positive $\muP$, leading to two distinct distributions.

\newcommand{\halffraction}{0.49}
\begin{figure}
  \centering
  \begin{subfigure}[t]{\halffraction\textwidth}
    \centering
    \includegraphics[width=\textwidth]{\imagepath/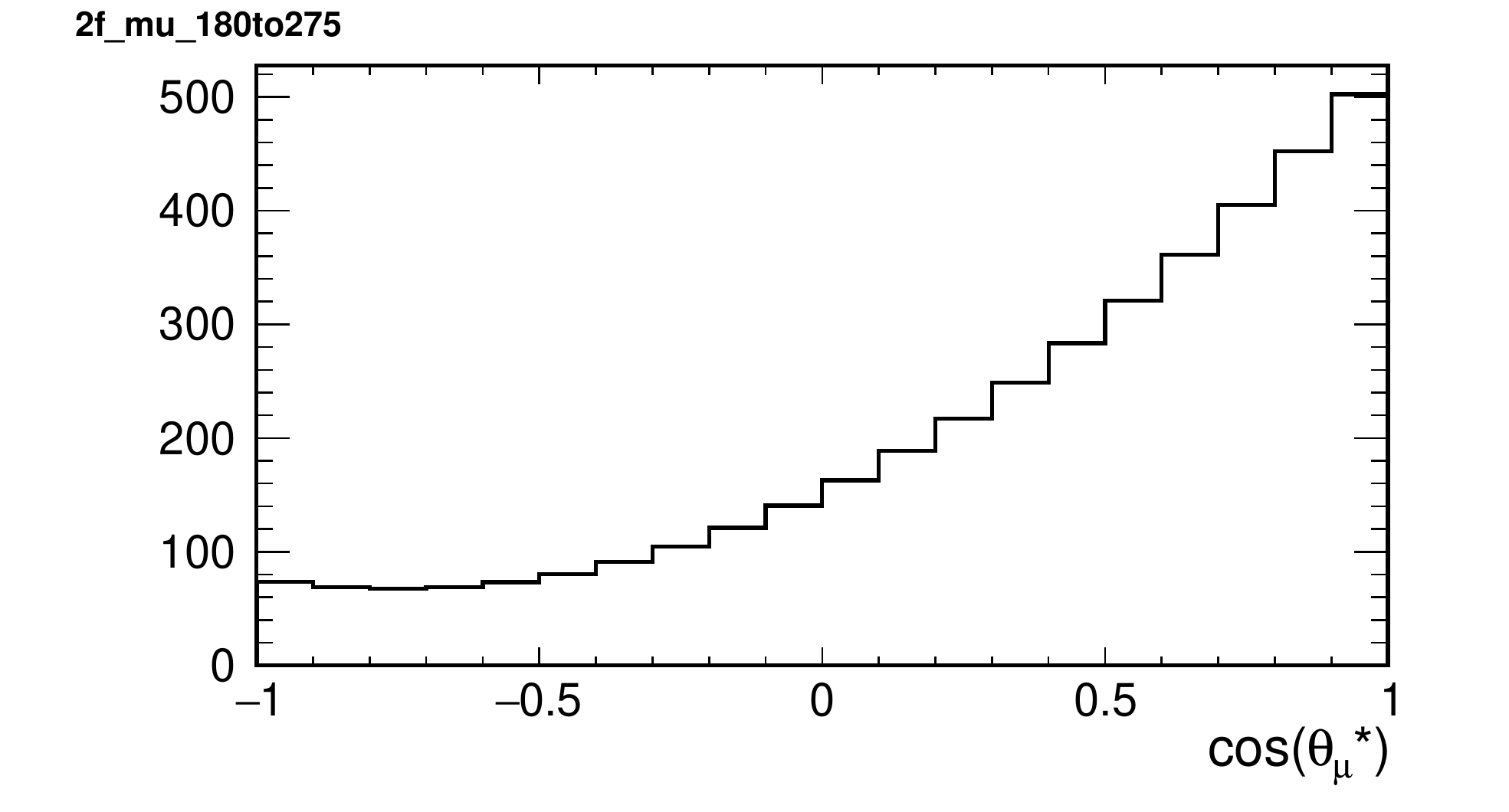}
    \caption{}\label{fig:DistributionExamples_mumu}
  \end{subfigure}
  \begin{subfigure}[t]{\halffraction\textwidth}
    \centering
    \includegraphics[width=\textwidth]{\imagepath/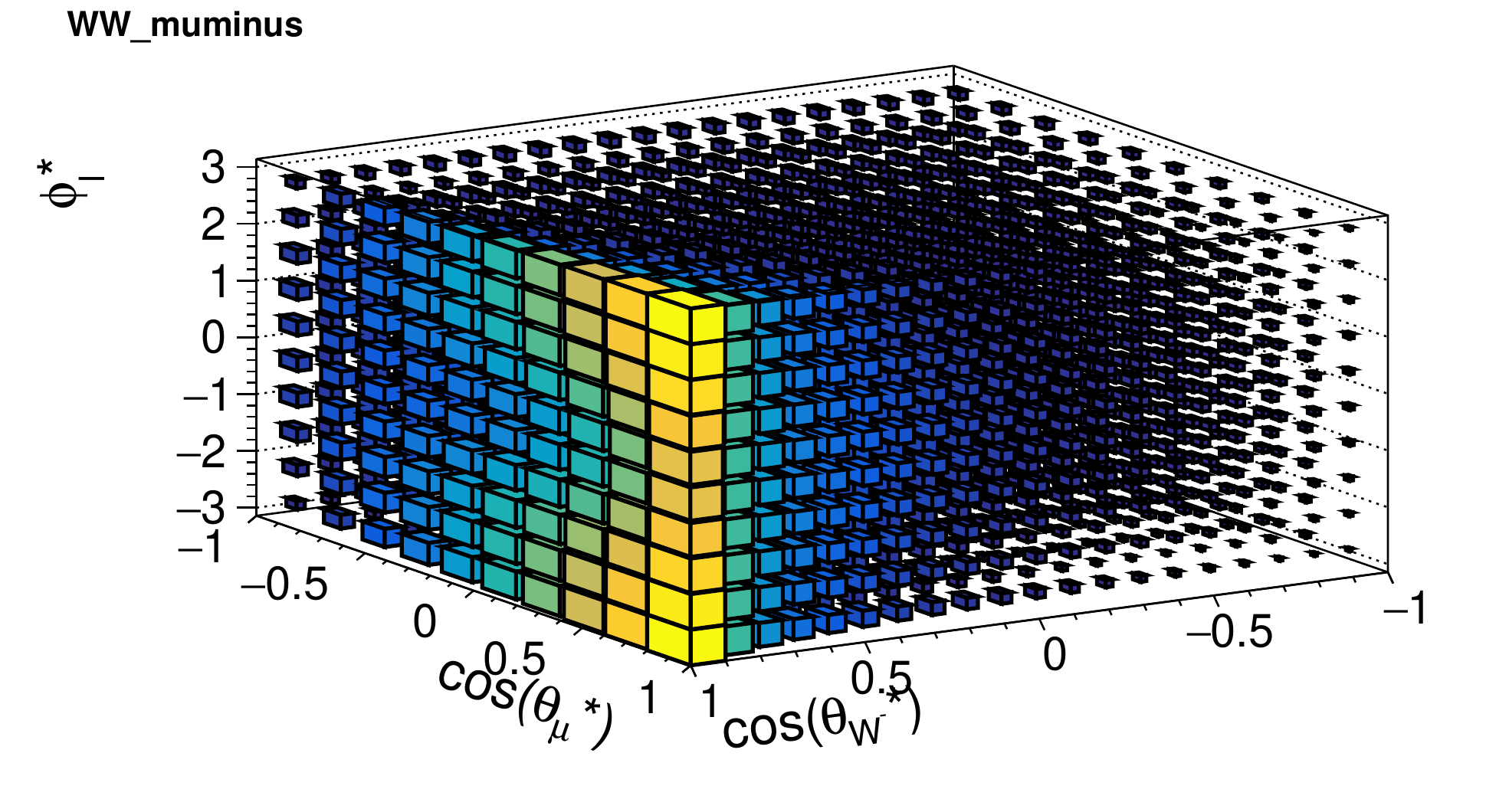}
    \caption{}\label{fig:DistributionExamples_4f}
  \end{subfigure}%
  \caption{%
    Generator-level differential distributions for%
     \subfigref{fig:DistributionExamples_mumu} $\mu\mu$ ($\sqrt{s*}\in[180,275]\,$GeV) and%
     \subfigref{fig:DistributionExamples_4f} $qq\muM\nubar\,(q=u,d,s,c,b)$ ($\sqrt{s*}\in[180,275]\,$GeV)%
    events for $\eML\ePR$ collisions at 250GeV from the ILD production \cite{Berggren_2021}.
  }
  \label{fig:DistributionExamples}
\end{figure}

\subsection{First test of the framework}

A first test of the framework uses the distributions and collider configurations described above to extract just the polarisation and luminosity parameters.
The test does not implement any other physical or systematic effects, and is a simple validation of the framework.
It assumes all physical constants and chiral cross sections, including chiral asymmetries, to be known exactly.

\begin{figure}
  \centering
  \includegraphics[width=0.55\textwidth]{\imagepath/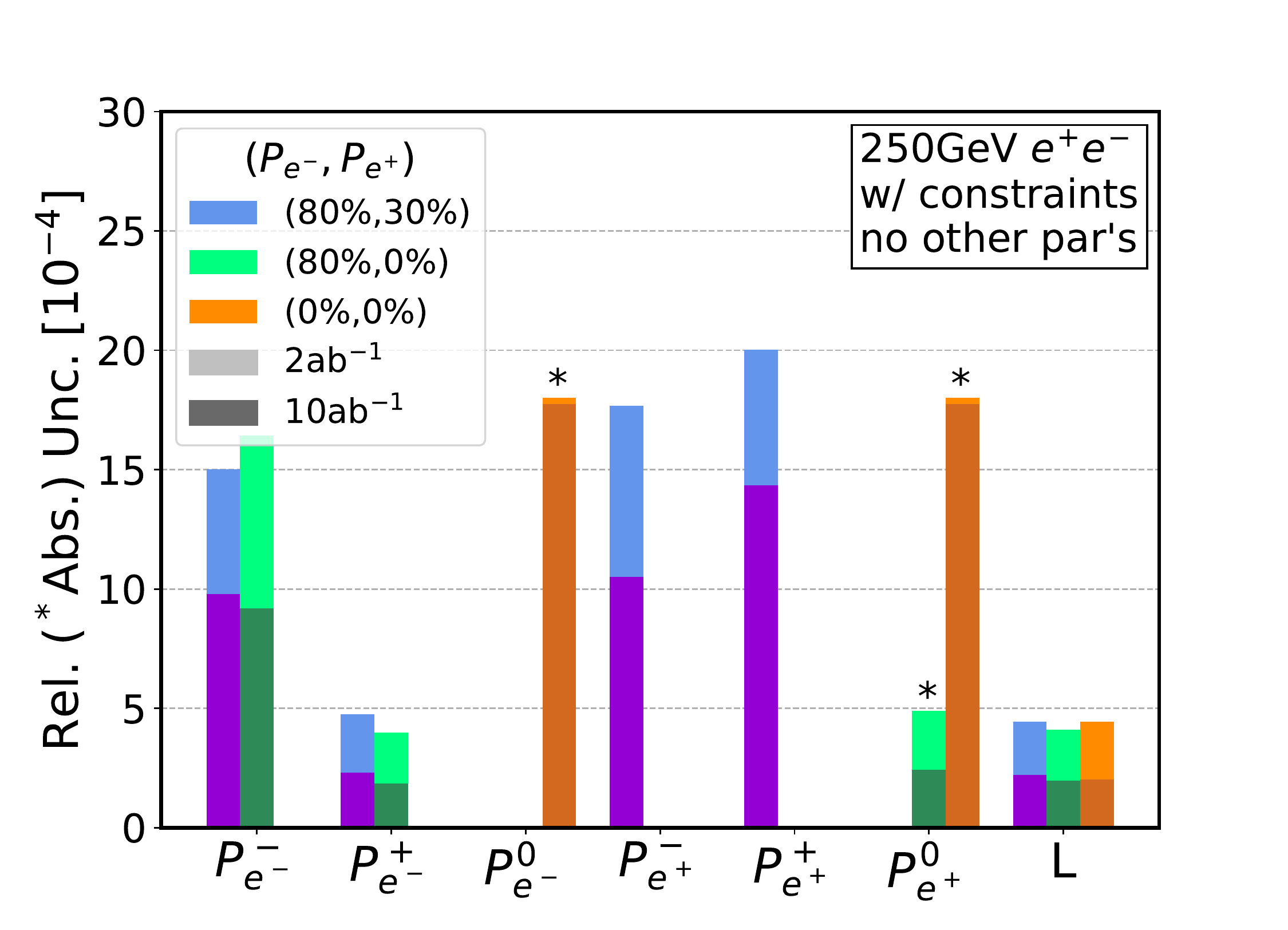}
  \caption{%
    Uncertainties for the luminosity and polarisation parameters for the different tested collider configurations in the first validation test of the framework.
  }
  \label{fig:UncPolLumiOnly}
\end{figure}

The uncertainties extracted from this test show an increased precision gained from higher luminosity (\cref{fig:UncPolLumiOnly}).
This gain does not always scale with the square-root of the luminosity increase because the uncertainty is a combination of the statistical uncertainty and the external constraint on the parameter.
In addition to the uncertainty on each parameter, the fit also provides the correlation matrix for each configuration (\cref{fig:CorrelationsPolLumiOnly}).
All correlations depend visibly on the collider scenario.
The unpolarised scenario does not allow an unambiguous determination of all 3 parameters from the distributions and relies on the external constraints.
Assuming the differential cross sections as perfectly known, all polarisations can be determined if at least one beam polarised.

This test lays the foundation for implementing further physical and systematic effects.

\newcommand{\thirdfraction}{0.33}
\begin{figure}
  \centering
  \begin{subfigure}[t]{\thirdfraction\textwidth}
    \centering
    \includegraphics[width=\textwidth]{\imagepath/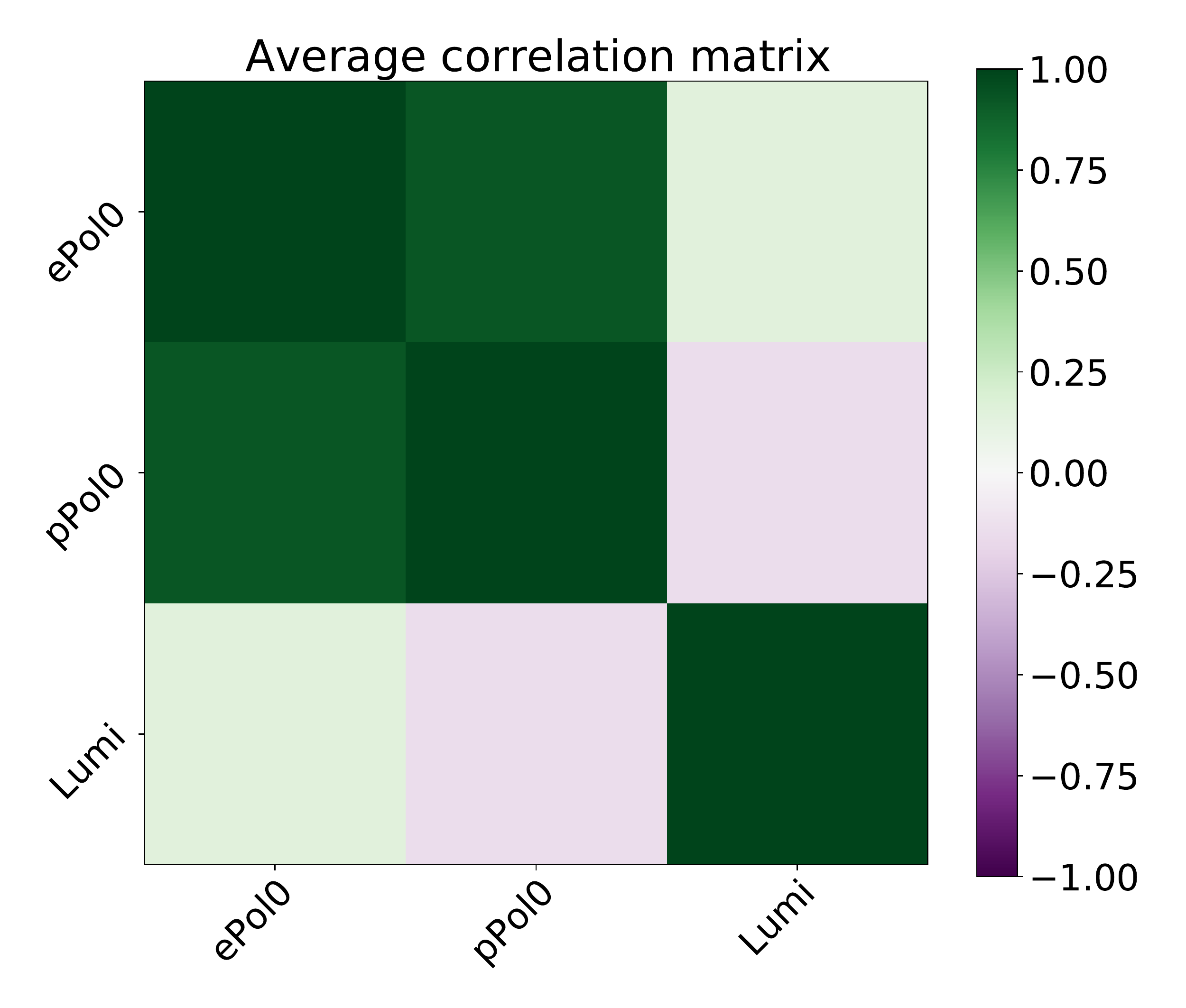}
  \end{subfigure}%
  \begin{subfigure}[t]{\thirdfraction\textwidth}
    \centering
    \includegraphics[width=\textwidth]{\imagepath/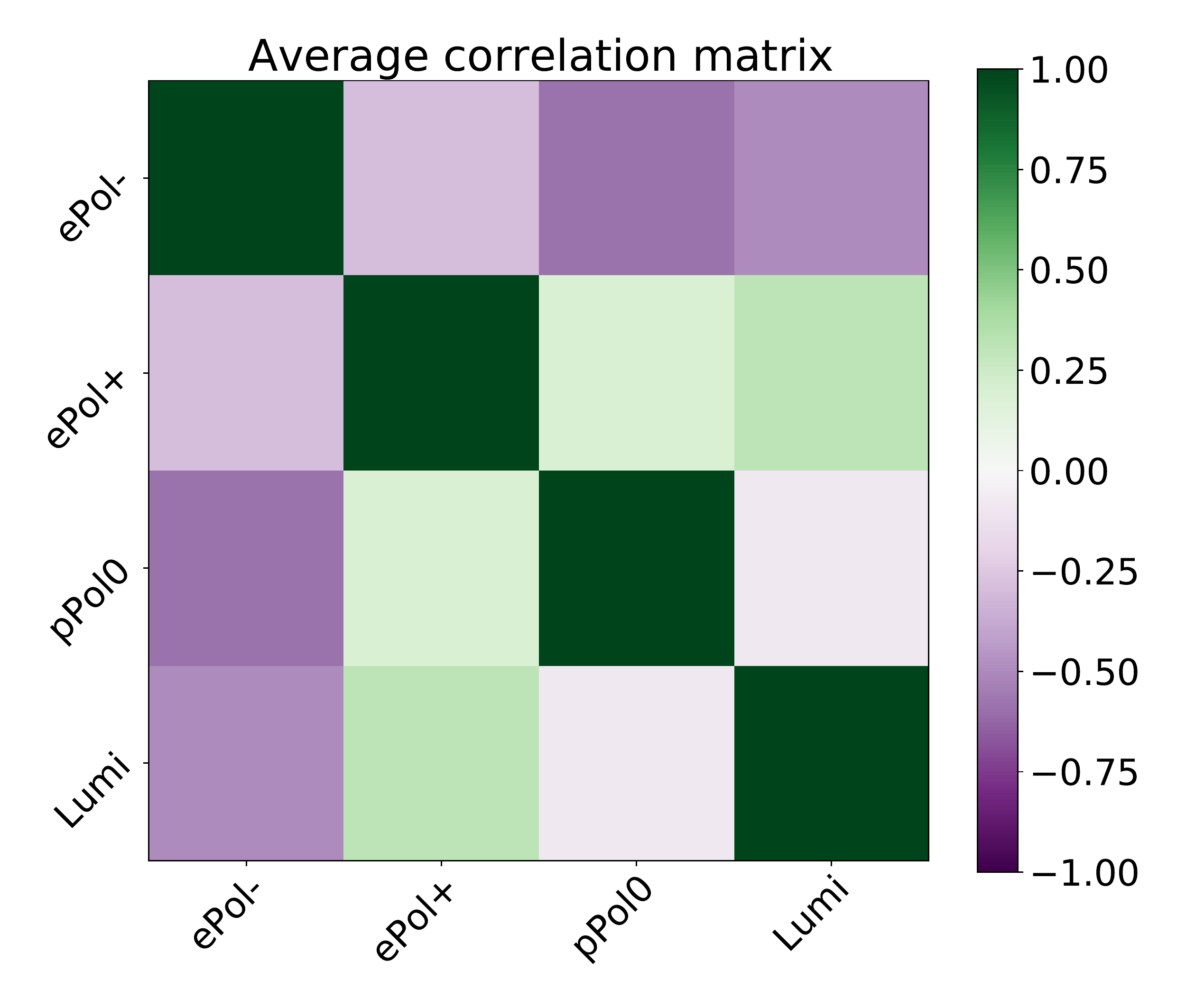}
  \end{subfigure}%
  \begin{subfigure}[t]{\thirdfraction\textwidth}
    \centering
    \includegraphics[width=\textwidth]{\imagepath/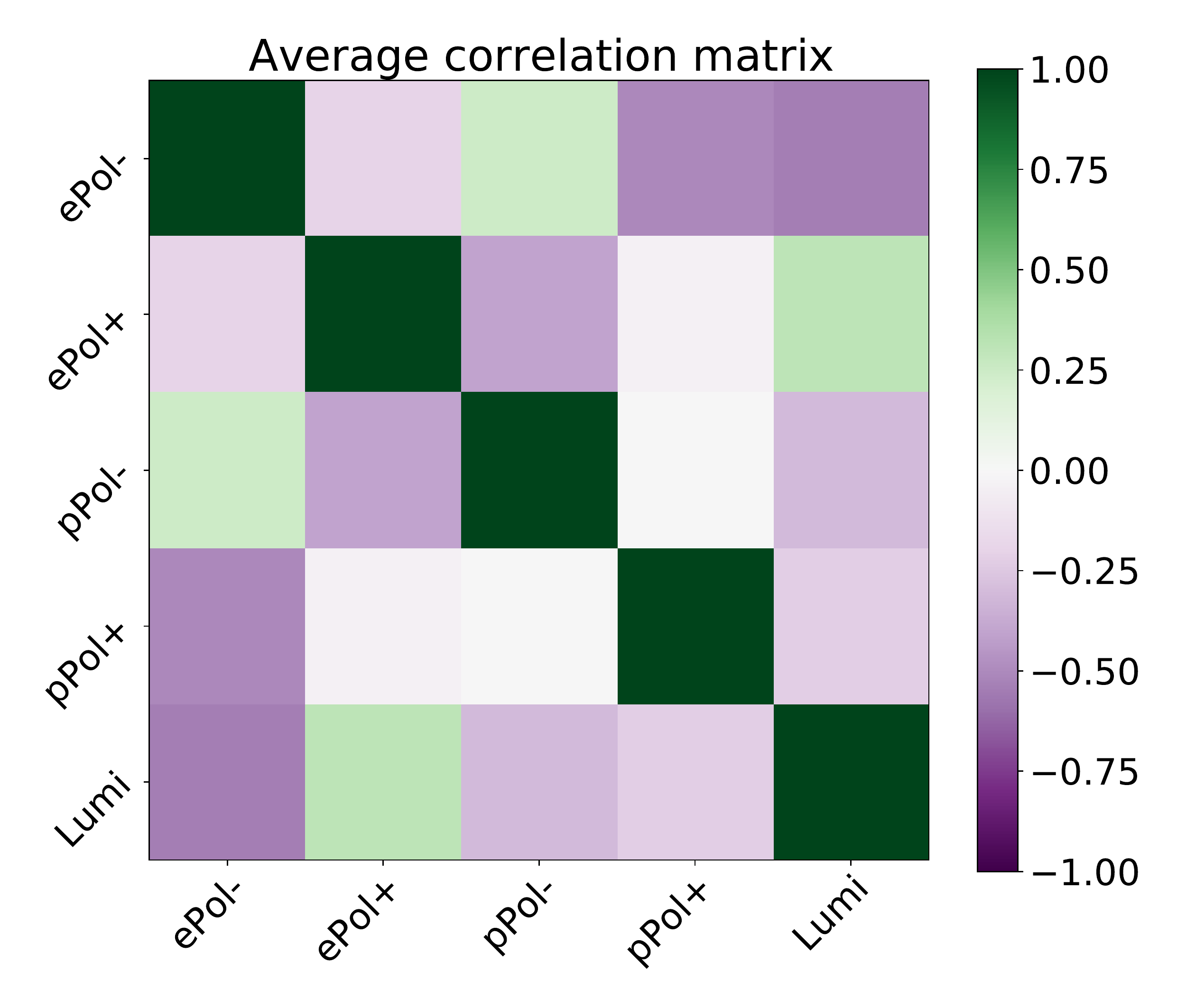}
  \end{subfigure}
  
  \begin{subfigure}[t]{\thirdfraction\textwidth}
    \centering
    \includegraphics[width=\textwidth]{\imagepath/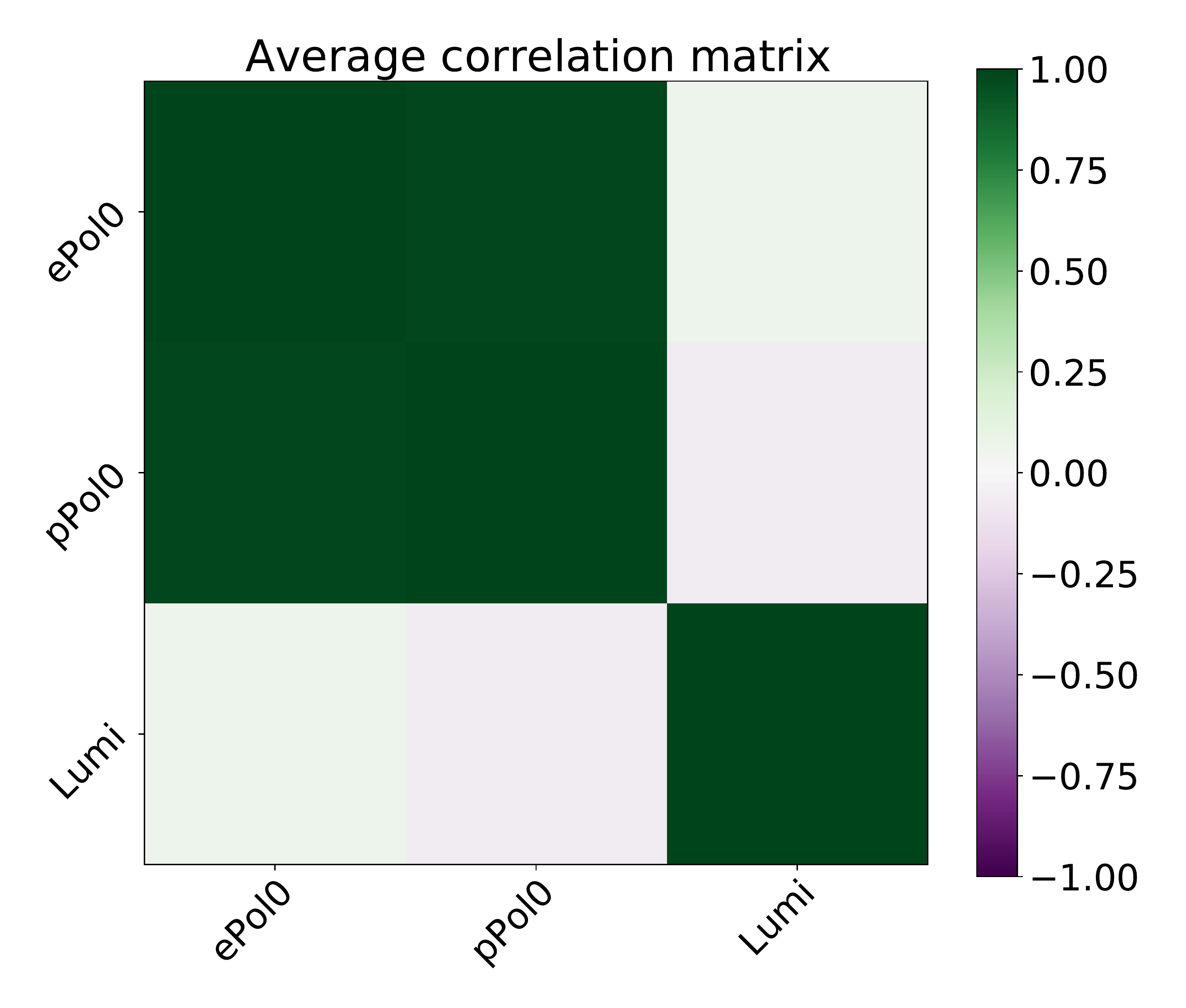}
    \end{subfigure}%
  \begin{subfigure}[t]{\thirdfraction\textwidth}
    \centering
    \includegraphics[width=\textwidth]{\imagepath/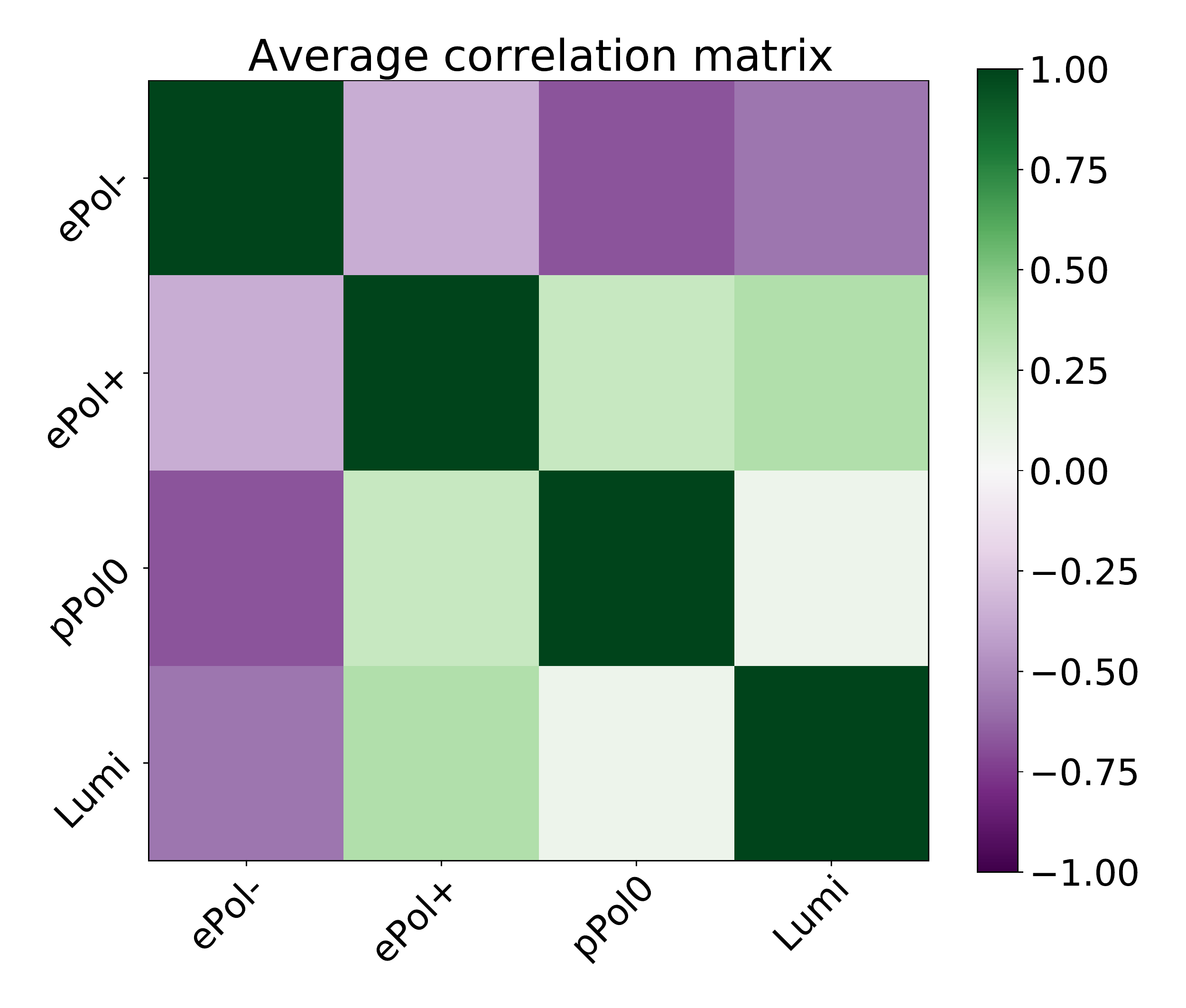}
  \end{subfigure}%
  \begin{subfigure}[t]{\thirdfraction\textwidth}
    \centering
    \includegraphics[width=\textwidth]{\imagepath/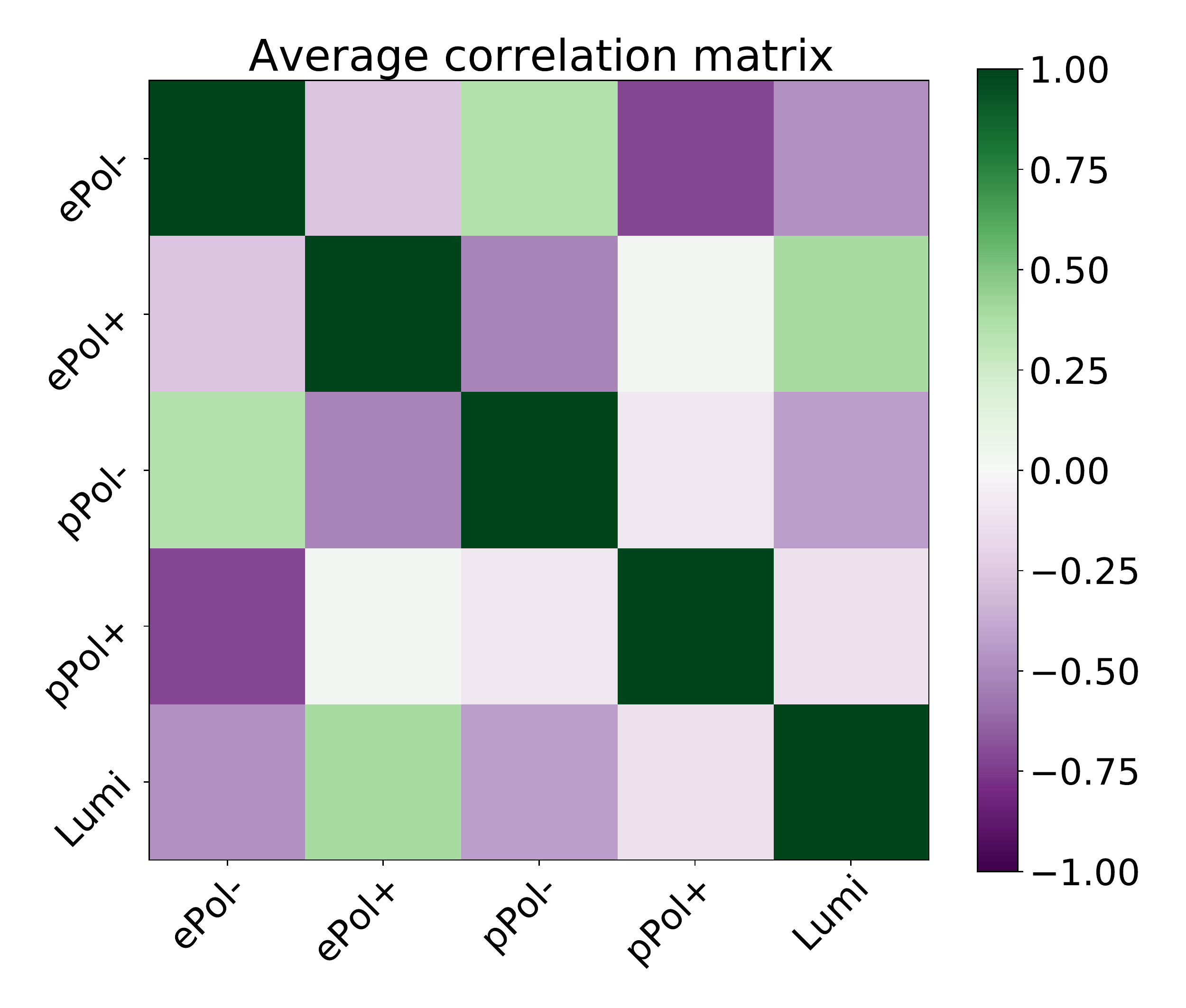}
  \end{subfigure}
  \caption{%
    Correlation matrices for the luminosity and polarisation parameters in the different collider configurations of the first validation test.
    The upper line uses $\Lumi=2\invab$, the lower $\Lumi=10\invab$.
    The three rows are from-left-to-right: unpolarised, electrons-only polarised, both beams polarised.
  }
  \label{fig:CorrelationsPolLumiOnly}
\end{figure}

\section{$\boldsymbol\mu$ acceptance parametrisation}
A test of the behaviour of a systematic effect in a combined fit requires an implementation of the effect into the fit.
This section describes the choice of the effect, the choice of the parametrisation in fit, and the validation of the parametrisation.

The choice of the systematic effect falls on the muon acceptance.
It is chosen because it has been a challenging uncertainty in previous electroweak measurements and it has a simple shape which is stable for high-momentum muons.
The tracking efficiency at future $\eP\eM$ detectors, which is the core component of the muon acceptance, is almost $100\%$ in the central region and largely independent of azimuthal angle, and sharply falls in very forward region (\cref{fig:MuAccPar_SiD}).

This behaviour corresponds approximately to an acceptance range with an upper edge $\costh_{up}$ and a lower edge $\costh_{low}$ of the muon polar angle in the detector frame, between which the acceptance is perfect.
A simple box acceptance model is therefore chosen as a first implementation, with a width $w = \costh_{up} - \costh_{low}$ and a center $c = \frac{1}{2} (\costh_{up} + \costh_{low})$ parameter (\cref{fig:MuAccPar_Box}).
An event passes this acceptance if all muons in the event are within the box range.
Here, the initial values for $c$ and $w$ correspond to edges at $\theta_{low} = 180\deg - \theta_{up} = 7\deg$.

The combined fit will extract these two parameters from the variations they cause on the differential distributions.
These variations need a parameterisation that can be implemented into the fit.
A quadratic parametrisation is chosen in terms of the variations $\left(\Dc=c'-c,\Dw=w'-w\right)$ of the parameters
  \begin{equation}
    \frac{d\sigma^{bin}}{\sigma^{bin}} = k_0^{bin} + k_c^{bin} \Dc + k_w^{bin} \Dw + k_{c2}^{bin} \Dc^2 + k_{w2}^{bin} \Dw^2 + k_{cw}^{bin} \Dc\Dw
  \end{equation}
where the coefficients $k$ are bin-dependent.

The coefficients of the polynomial are determined by applying the box-shaped acceptance on the generator-level events and slightly varying the cut parameters.
A common scale $\delta$ of these variations describes the order-of-magnitude of the expected uncertainties.
Here, $\delta = 2\times10^{-4}$ is chosen.
The coefficient determination tests 29 points $(c',w')$ in the range $\sqrt{\Dc^2 + \Dw^2} \in [0,1.5]\delta$.
It first chooses 6 points and solves the linear equation system to get an initial estimate of the coefficients.
Then, using these starting values, it fits the polynomial to all 29 points and extracts the final coefficients.

The parametrisation step requires a validation.
Such a validation needs to show that the quadratic polynomial approach accurately estimates the box-shaped acceptance on all affected bins.
A definition and comparison of the two following quantities for each tested $\left(\Dc,\Dw\right)$ point can provide such a validation.
The first quantity evaluates the impact of changing the cut parameters when applying the cut on the events.
\begin{equation}
  \chisqshift = \sum_{\text{affected bins}} \left(\frac{N_{cut}^{(\Delta c, \Delta w)} - N_{cut}^{0}}{\sqrt{N_{cut}^{0}}}\right)^2
\end{equation}
$N_{cut}$ describes the number of events remaining after the cut, and $N_{cut}^{0}$ uses the initial non-varied $c$ and $w$ values.
The second quantity tests the impact of the deviation that parametrisation shows with respect to the box-shaped cut applied to the events.
\begin{equation}
  \chisqpar = \sum_{\text{affected bins}} \left(\frac{N_{par}^{(\Delta c, \Delta w)} - N_{cut}^{(\Delta c, \Delta w)}}{\sqrt{N_{cut}^{(\Delta c, \Delta w)}}}\right)^2
\end{equation}
$N_{par}$ is the number of events that the parametrisation yields.

\begin{figure}
  \centering
  \includegraphics[width=0.65\textwidth]{\imagepath/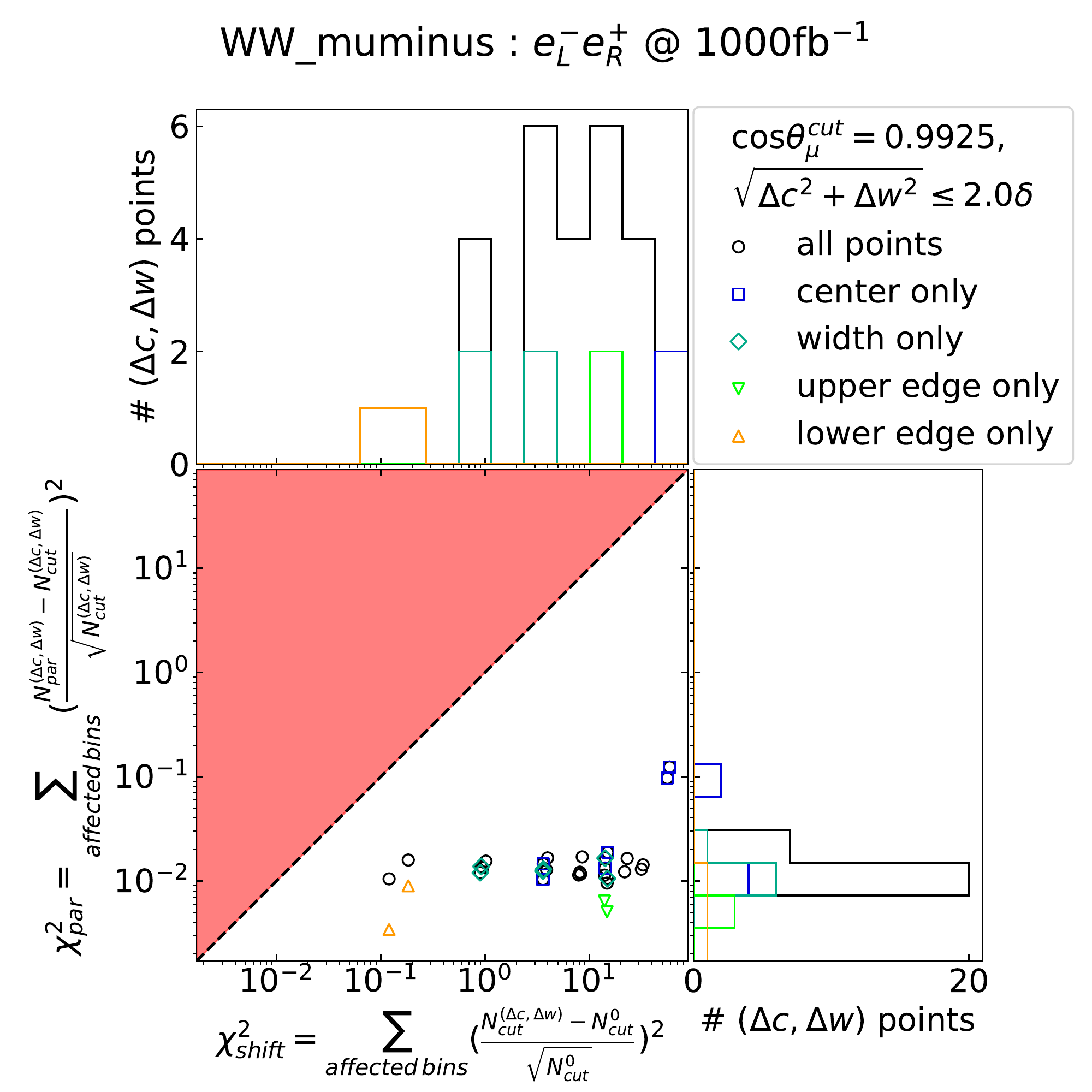}
  \caption{%
    Example for the validation test of the muon acceptance parametrisation that is employed in the fit, using $qq\muM\nubar\,(q=u,d,s,c,b)$ ($\sqrt{s*}\in[180,275]\,$GeV) events for $\eML\ePR$.
  }
  \label{fig:WW_chisqtest}
\end{figure}

The validation tests many different combinations of $\Dc$ and $\Dw$ in range of $\sqrt{\Dc^2 + \Dw^2} \in [0.5,2]\delta$.
At each point, the impact of cut can be considered more relevant than the deviation caused by the parametrisation if the tested point has $\chisqshift > \chisqpar$.
Most tested points are found to be far in $\chisqshift \gg \chisqpar$ region, indicating that the parametrisation is safe to use (\cref{fig:WW_chisqtest}).
For the $\mu\mu$ distributions, a few points are found at the $\chisqshift \approx \chisqpar$ line.
These are unproblematic because they have a low impact on the distribution and other distributions strongly restrict the corresponding direction.

\renewcommand{\halffraction}{0.49}
\begin{figure}
  \centering
  \begin{subfigure}[t]{\halffraction\textwidth}
    \centering
    \includegraphics[width=\textwidth]{\imagepath/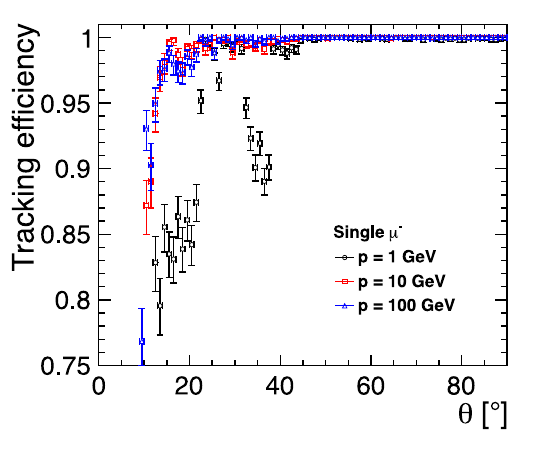}
    \caption{}\label{fig:MuAccPar_SiD}
  \end{subfigure}
  \begin{subfigure}[t]{\halffraction\textwidth}
    \centering
    \includegraphics[width=\textwidth]{\imagepath/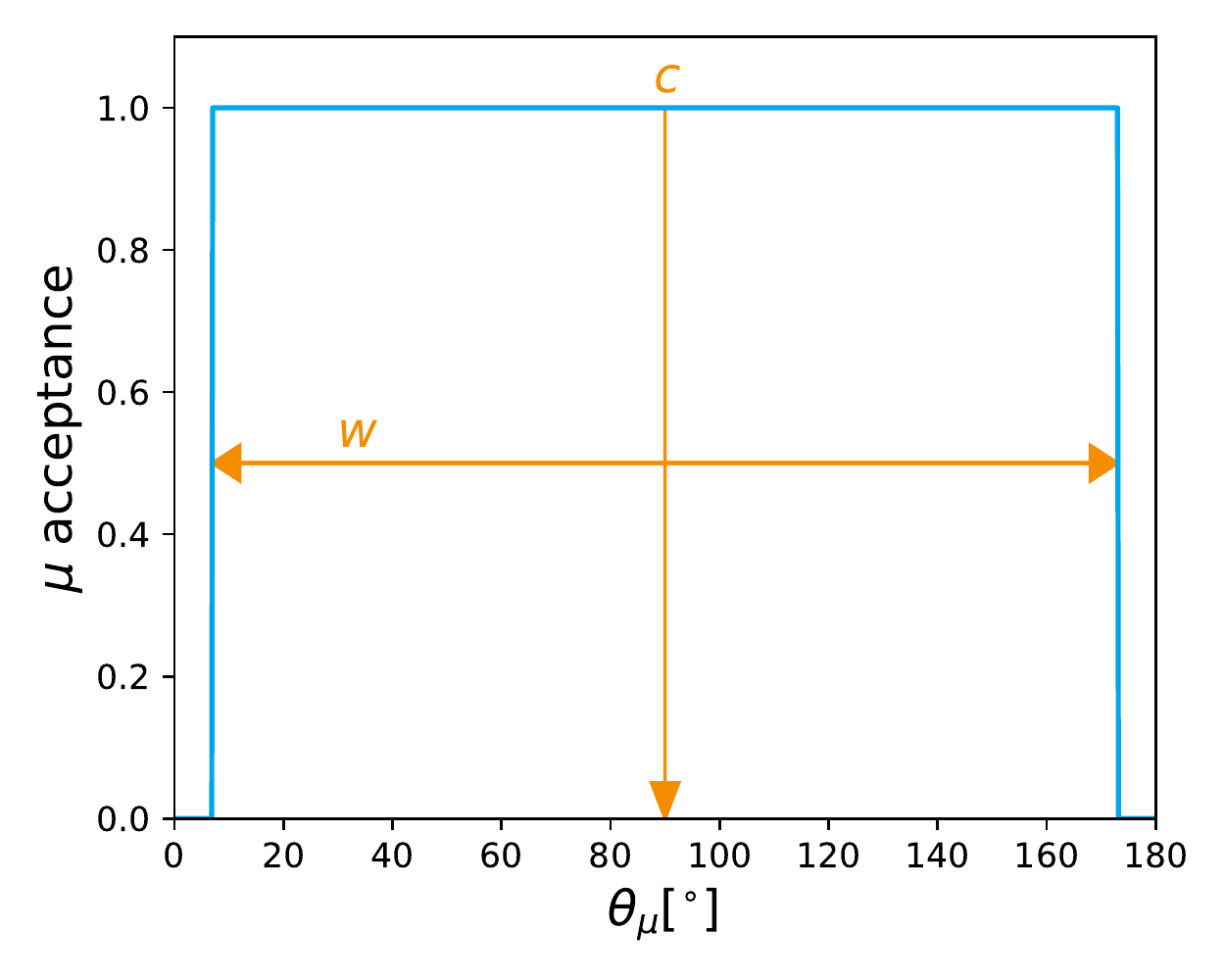}
    \caption{}\label{fig:MuAccPar_Box}
  \end{subfigure}%
  \caption{%
    Comparison of \subfigref{fig:MuAccPar_SiD} a realistic tracking efficiency \cite{behnke2013international} and \subfigref{fig:MuAccPar_Box} the simple box-shape model for the muon acceptance which is used in this study.
  }
  \label{fig:MuAccPar}
\end{figure}

\section{$\boldsymbol\mu$ acceptance in different collider setups}
This section presents a first test of the implementation of the box-shaped muon acceptance in the different collider configurations described above.
Each fit considers the corresponding luminosity and polarisation parameters as well as the muon acceptance parameters.
It does not consider any physical effects, assumes all chiral cross sections to be known exactly, and does not include backgrounds or high-level selection efficiencies.

\begin{figure}
  \centering
  \includegraphics[width=0.55\textwidth]{\imagepath/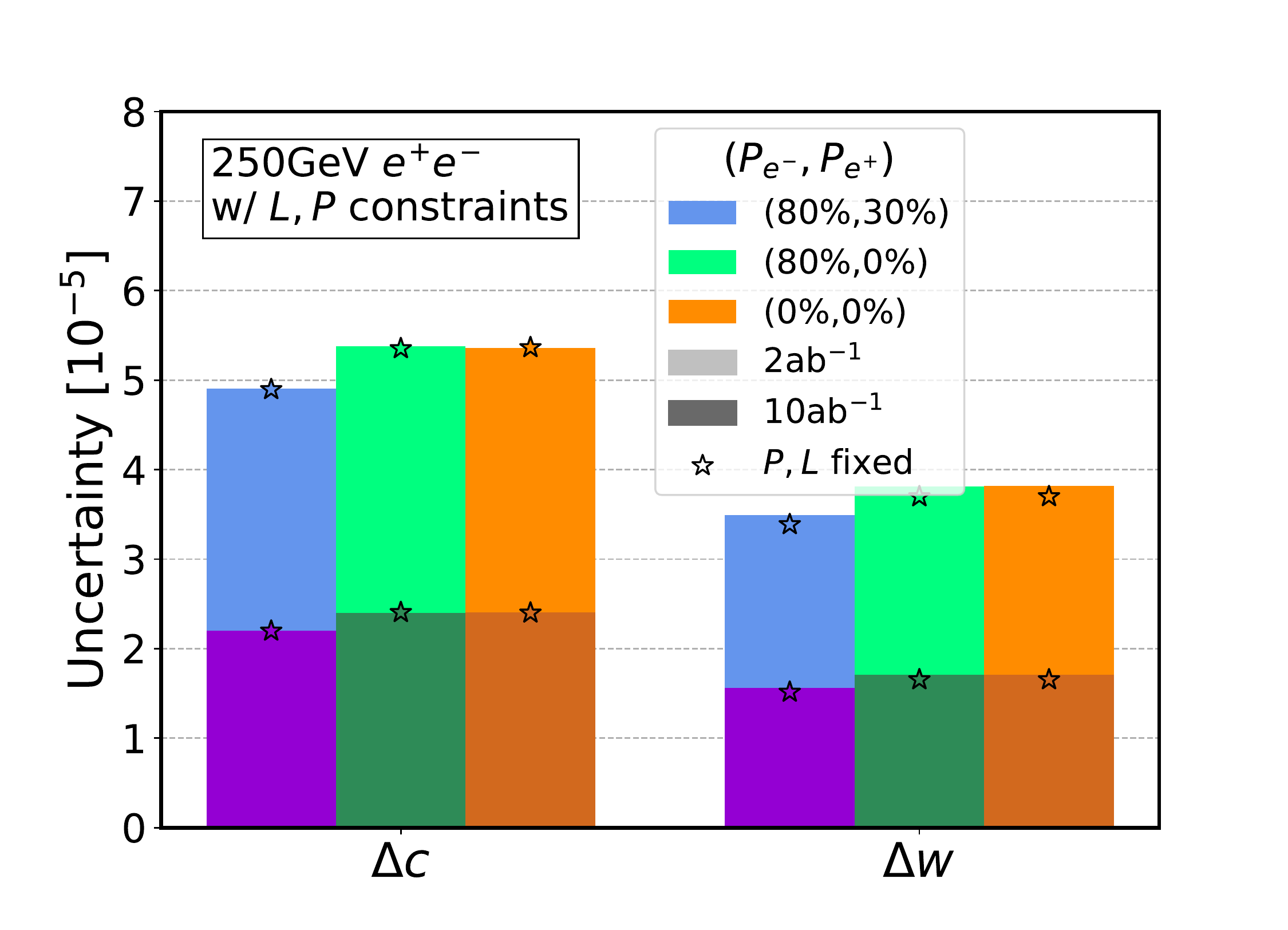}
  \caption{%
    Uncertainties for the muon acceptance parameters for the different tested collider configurations in the test that involved the muon acceptance and the luminosity and polarisation parameters.
  }
  \label{fig:UncMuAcc}
\end{figure}

Under these idealised conditions, the fits determine the $c$ and $w$ parameters with an accuracy of the order of $10^{-5}$ (\cref{fig:UncMuAcc}).
Observed differences come from changes in the statistics, either a factor $\sqrt{5}$ from the luminosity change, or the optimised luminosity sharing in the case of where both beams are polarised.
Fixing the luminosity and polarisation parameters leaves the results unchanged.

The acceptance parameters do not correlate with each other or with the polarisations (\cref{fig:CorrelationsWithMuAcc}).
A slight anticorrelation of the width parameter $w$ with the luminosity originates from the common effect of increasing the integrated number of events.
Changes in the collider configuration do not influence the correlations of the muon acceptance parameters.

This first test for the inclusion of systematic effects does not include any physical effects.
It cannot make any statement on the potential advantages of beam polarisation with respect to systematic uncertainties.
Such statements will require the implementation of physical effects, which will be the next steps in this study.

\renewcommand{\thirdfraction}{0.33}
\begin{figure}
  \centering
  \begin{subfigure}[t]{\thirdfraction\textwidth}
    \centering
    \includegraphics[width=\textwidth]{\imagepath/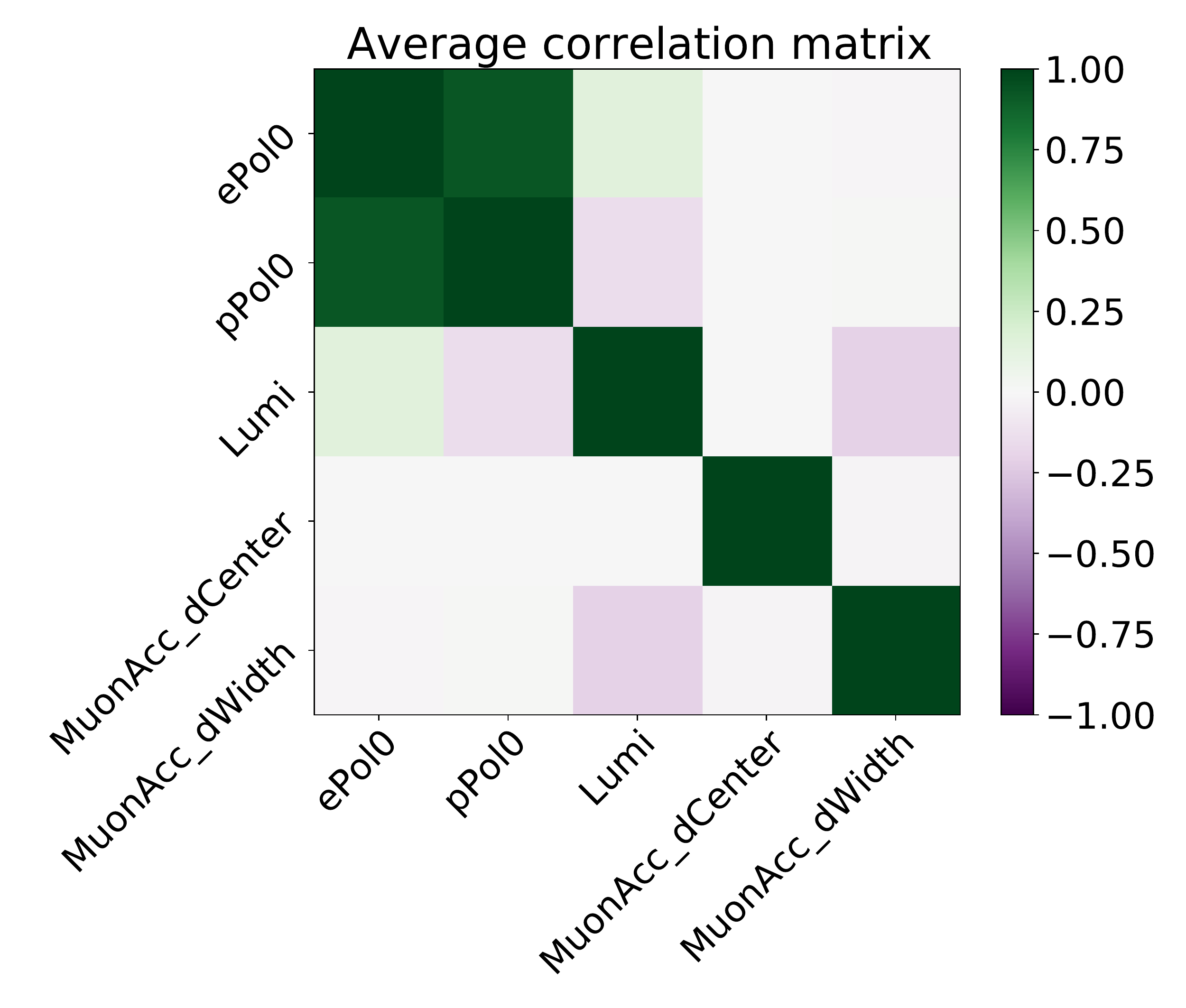}
  \end{subfigure}%
  \begin{subfigure}[t]{\thirdfraction\textwidth}
    \centering
    \includegraphics[width=\textwidth]{\imagepath/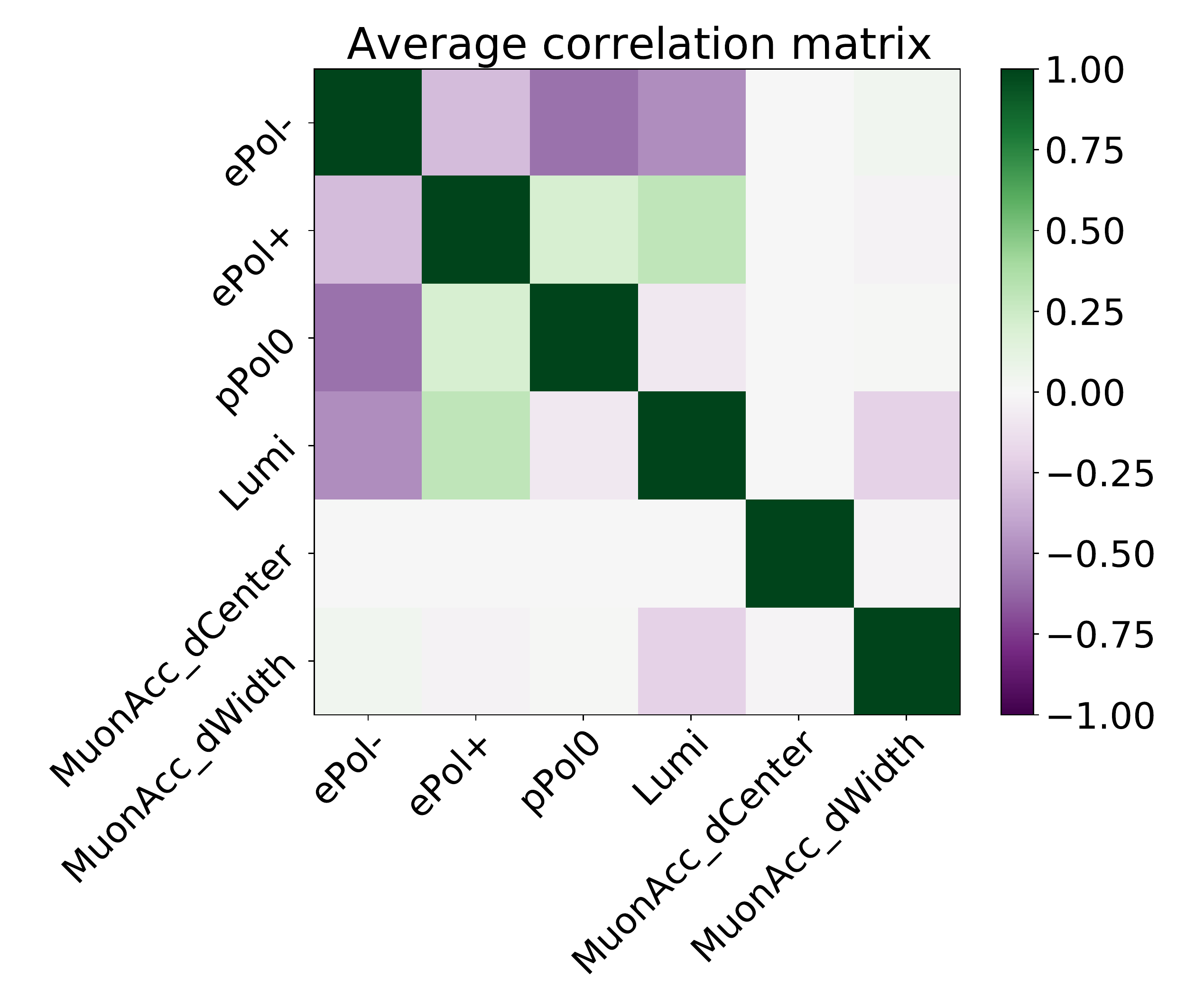}
  \end{subfigure}%
  \begin{subfigure}[t]{\thirdfraction\textwidth}
    \centering
    \includegraphics[width=\textwidth]{\imagepath/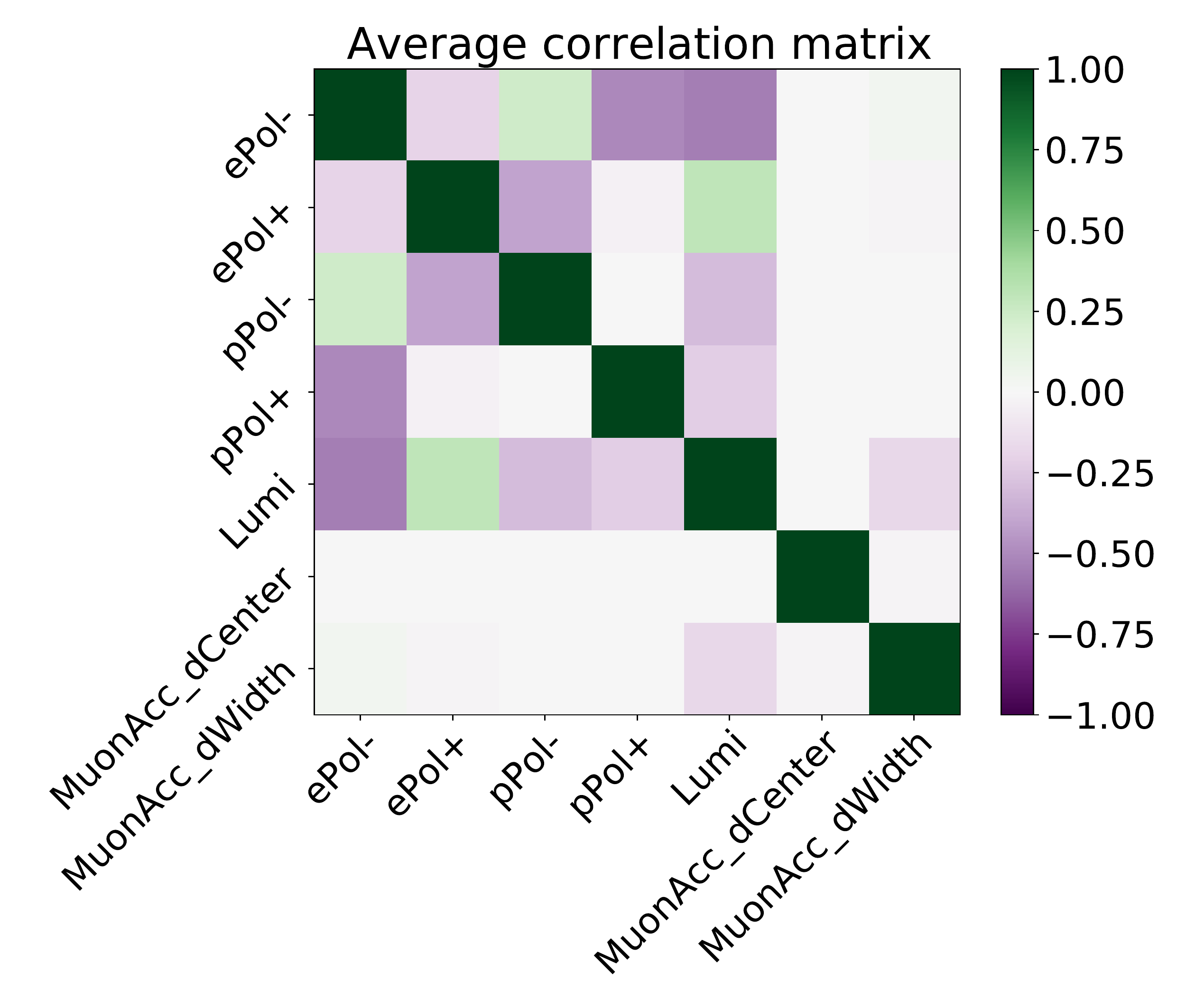}
  \end{subfigure}
  
  \begin{subfigure}[t]{\thirdfraction\textwidth}
    \centering
    \includegraphics[width=\textwidth]{\imagepath/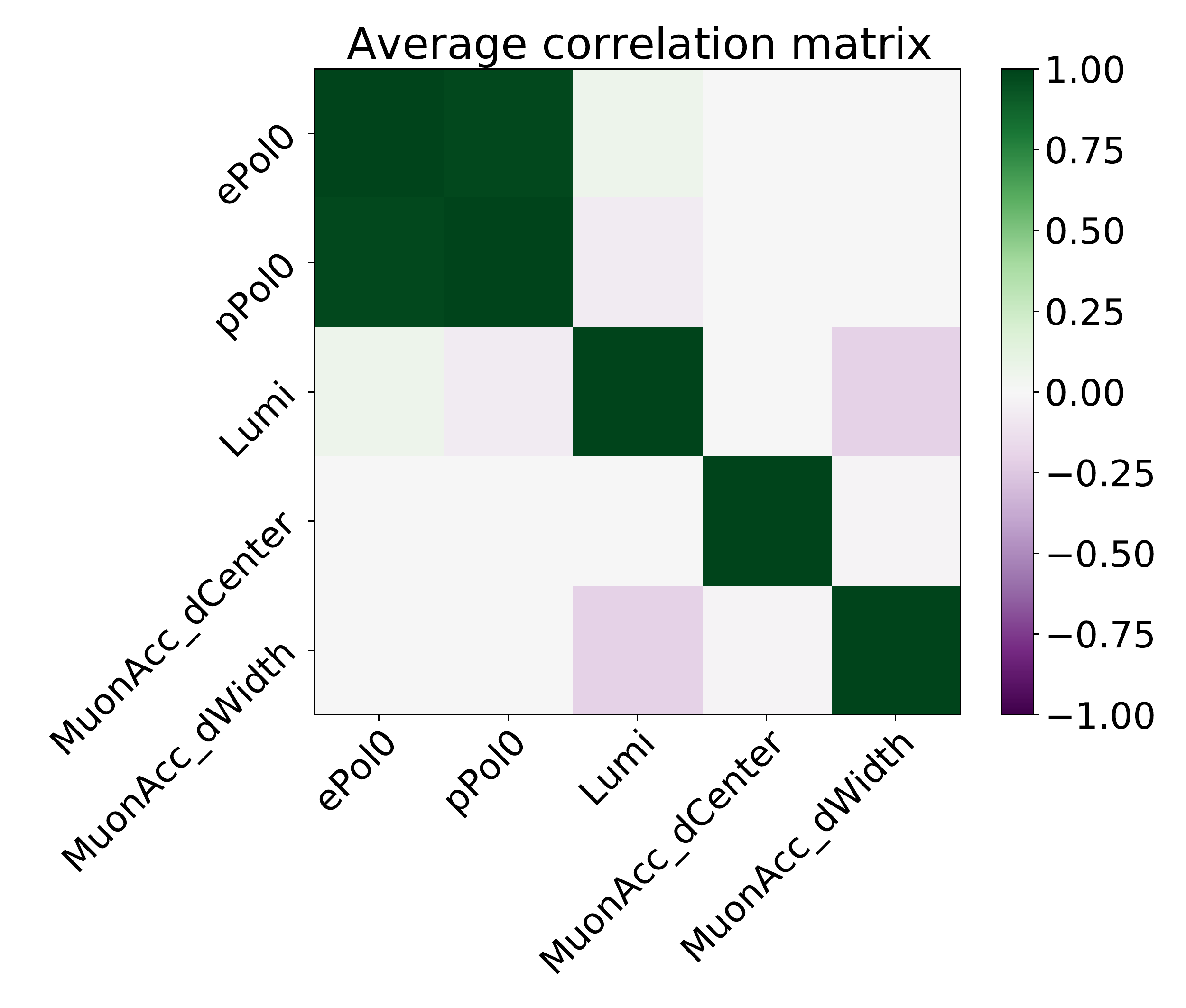}
    \end{subfigure}%
  \begin{subfigure}[t]{\thirdfraction\textwidth}
    \centering
    \includegraphics[width=\textwidth]{\imagepath/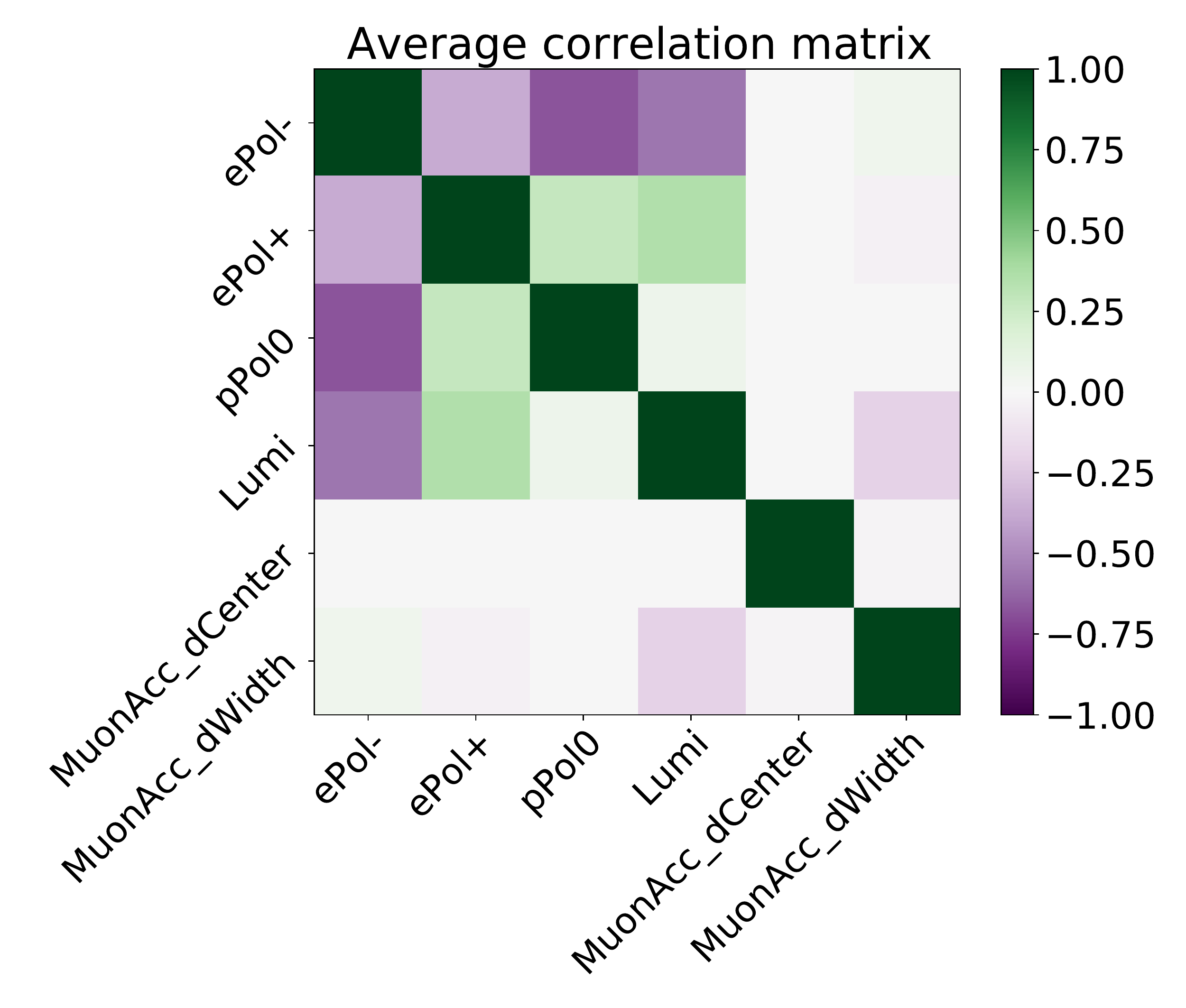}
  \end{subfigure}%
  \begin{subfigure}[t]{\thirdfraction\textwidth}
    \centering
    \includegraphics[width=\textwidth]{\imagepath/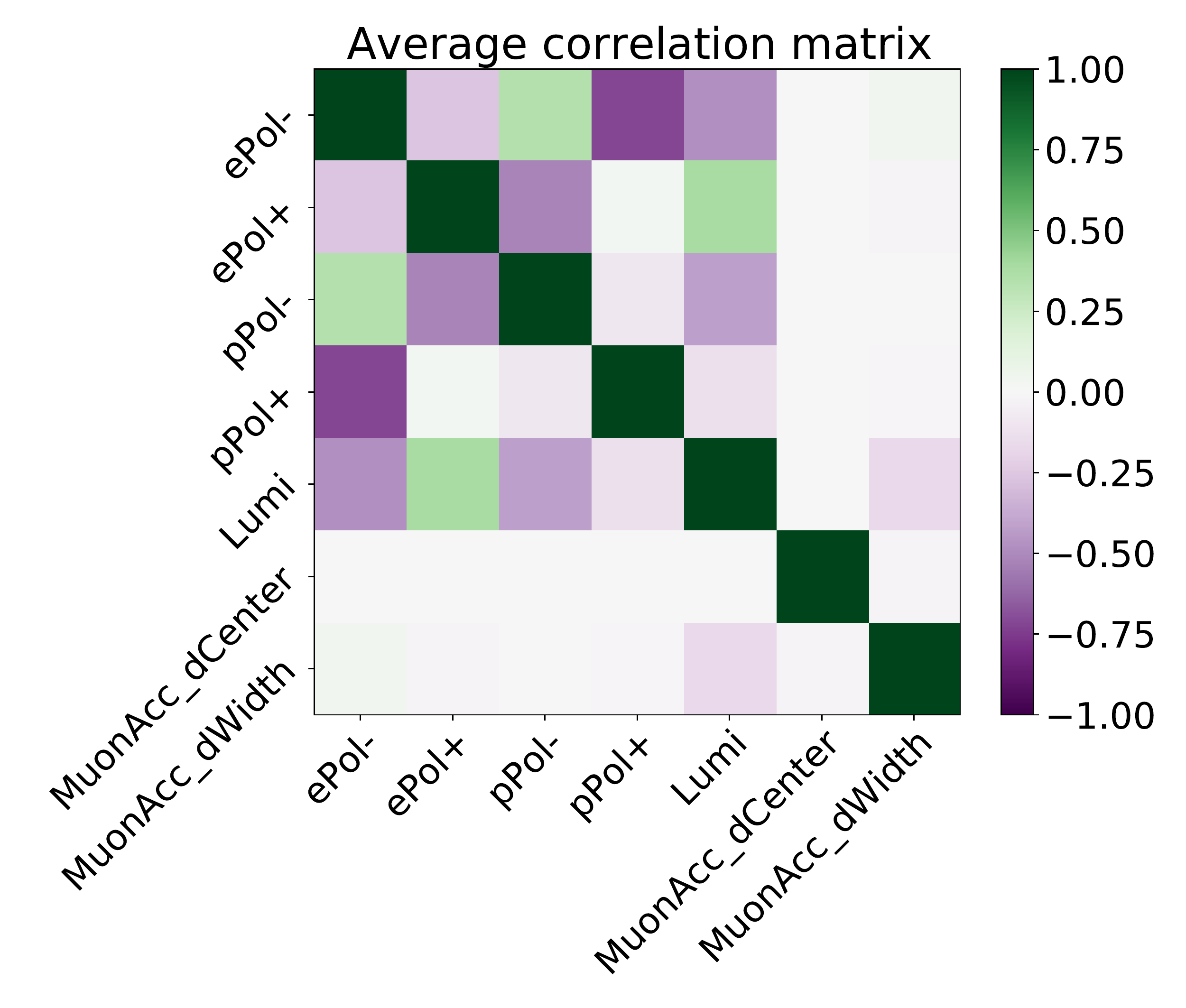}
  \end{subfigure}
  \caption{%
    Correlation matrices for the parameters in the test that involved the muon acceptance, the luminosity and polarisation.
    The upper line uses $\Lumi=2\invab$, the lower $\Lumi=10\invab$.
    The three rows are from-left-to-right: unpolarised, electrons-only polarised, both beams polarised.
  }
  \label{fig:CorrelationsWithMuAcc}
\end{figure}

\section{Conclusion \& Outlook}
Beam polarisation may help reduce the impact of systematic uncertainties when all different polarised datasets are  included in a combined fit.
A fit framework is set up that can perform combined fits on the different datasets of a polarised collider.
First tests in different collider configurations validate the usage of the framework for the case of extracting the beam polarisation and luminosity parameters from $\mu\mu$ and $qq\mu\nu$ differential distributions.
The implementation and validation of a simplified muon acceptance into the fit shows that it may be feasible to directly include realistic systematic effects. 
Further studies that include physical effects will allow tests of the impact of beam polarisation on systematic uncertainties.

\subsection*{Acknowledgments}
This work was funded by the Deutsche Forschungsgemeinschaft under Germany’s Excellence Strategy – EXC 2121 ``Quantum Universe'' – 390833306.
It has benefited from the computing services provided by the German National Analysis Facility (NAF)\cite{Haupt_2010}.

\section*{References}
\customprintbibliography

\end{document}